\documentclass{article}
\usepackage{spconf,amsmath,graphicx}
\usepackage{hyperref}
\usepackage{xspace}
\usepackage{xcolor}
\usepackage{subcaption}
\usepackage{graphicx}
\usepackage{booktabs}

\newcommand{\dataset}{\texttt{Visual-ASR-EC}\xspace}

\usepackage{multirow}

\title{Visual Information Matters for ASR Error Correction}
\name{Vanya Bannihatti Kumar, Shanbo Cheng, Ningxin Peng, Yuchen Zhang}
\address{ByteDance \\
\{vanya.bk, chengshanbo, nxpeng, zhangyuchen.zyc\}@bytedance.com
}
\begin{document}
%
\maketitle
\begin{abstract}

Aiming to improve the Automatic Speech Recognition (ASR) outputs with a post-processing step, ASR error correction (EC) techniques have been widely developed due to their efficiency in using parallel text data. Previous works mainly focus on using text or/ and speech data, which hinders the performance gain when not only text and speech information, but other modalities, such as visual information are critical for EC. The challenges are mainly two folds: one is that previous work fails to emphasize visual information, thus rare exploration has been studied. The other is that the community lacks a high-quality benchmark where visual information matters for the EC models. Therefore, this paper provides 1) simple yet effective methods, namely gated fusion and image captions as prompts to incorporate visual information to help EC; 2) large-scale benchmark dataset, namely \dataset\footnote{\url{https://github.com/VanyaBK/visual_ASR_EC}}, where each item in the training data consists of visual, speech, and text information, and the test data are carefully selected by human annotators to ensure that even humans could make mistakes when visual information is missing. Experimental results show that using captions as prompts could effectively use the visual information and surpass state-of-the-art methods by upto 1.2\% in Word Error Rate(WER), which also indicates that visual information is critical in our proposed \dataset dataset.

%

\end{abstract}
\section{Introduction}
\label{sec:intro}



\begin{figure*}[htb!]
\centering
\begin{subfigure}{0.4\textwidth}
  \centering
  \includegraphics[width=0.8\linewidth]{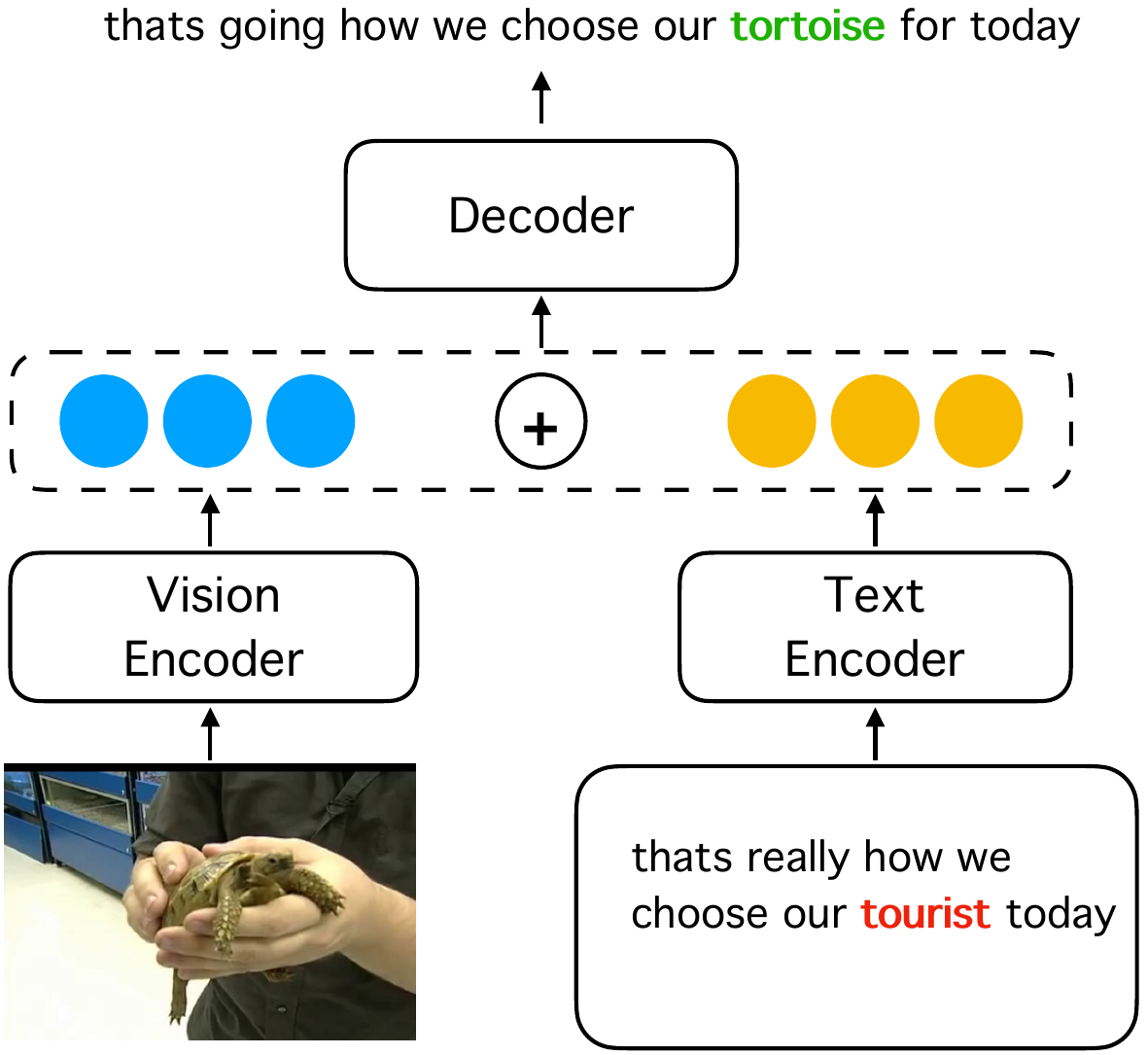}  
  \caption{Fusion-based}
  \label{fig:fusion}
\end{subfigure}
\begin{subfigure}{.4\textwidth}
  \centering
  \includegraphics[width=0.8\linewidth]{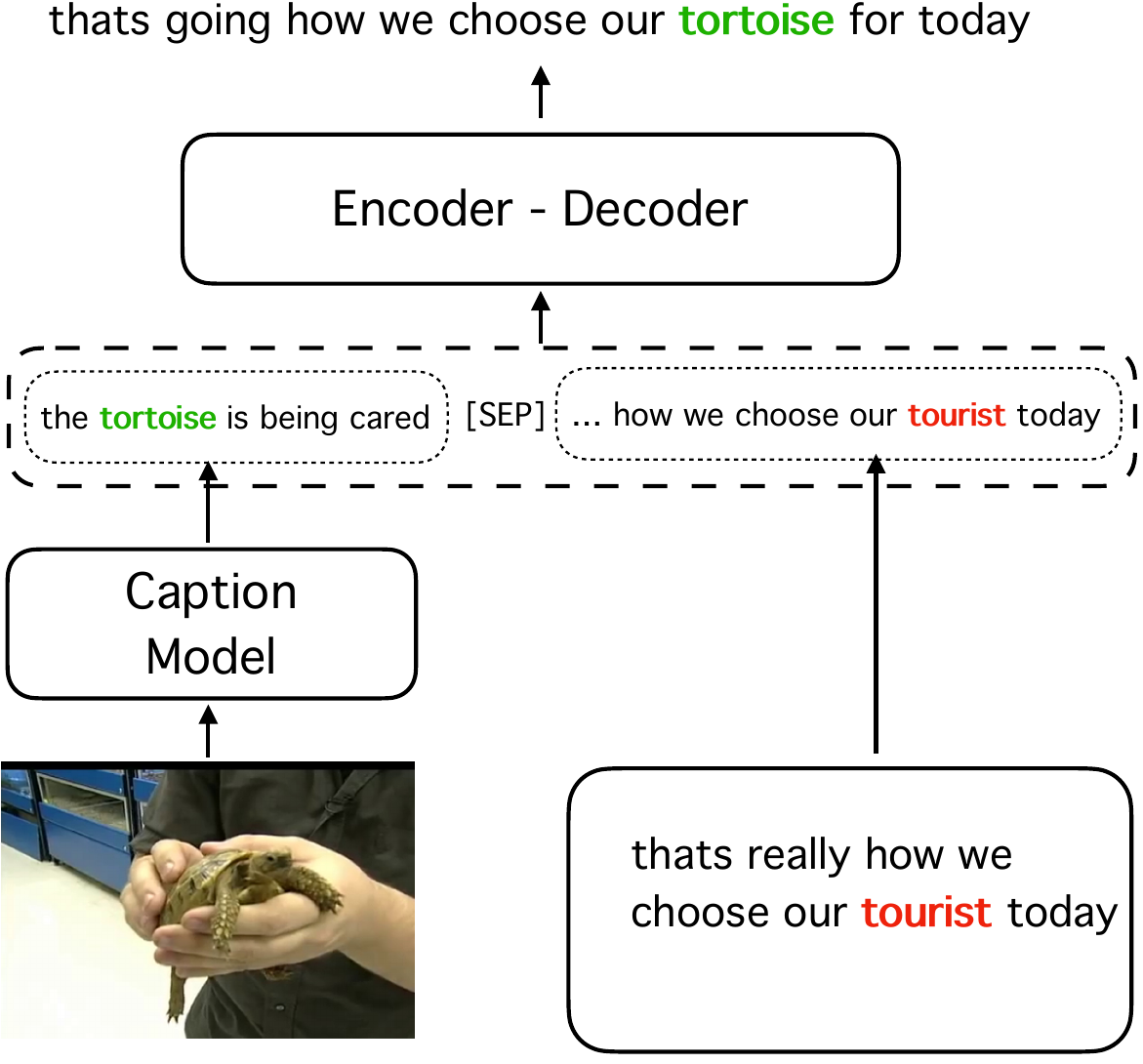}  
  \caption{Prompt-based}
  \label{fig:prompt}
\end{subfigure}
\caption{Illustration of the proposed methods for Visual ASR Correction. For the gated fusion method(left), the visual features and text embeddings are concatenated and sent to a decoder to get the corrected output. For the prompt-based method(right), given a source sentence to be corrected, we first generate the corresponding image caption as additional context to the source input, and then use a sequence-to-sequence model to generate the final result.}
\label{fig:caption}
\end{figure*}

Over the past years, automatic speech recognition (ASR) models have achieved great success \cite{NEURIPS2021_ea159dc9,46687}. However, there are still many errors in the ASR outputs caused by inherent difficulties, such as grammatical incoherence, homophone errors, etc.



To alleviate grammatical errors, some previous work propose to use an error correction module, typically sequence-to-sequence models, to correct the ASR outputs, with the help of large-scale text data\cite{inproceedings}. But since the error patterns vary irregularly based on context, pronunciation and language understanding, it is challenging to construct good quality of ASR hypotheses to reference text pairs to be used for supervised training. As the homophone errors are more difficult and might not be corrected with only text information, other works propose to utilize audio information in terms of phonetics for correction \cite{inproceedings_sc,9746763}. Although great works have been studied, there are still certain errors that cannot be properly corrected with audio and text information. For example, the sentence "\texttt{if you plan a little bit}" and "\texttt{if you plant a little bit}" sounds quite similar to each other, and both are grammatically correct. In such cases, equipping the correction models with only audio and text information is not sufficient, and other modalities, for example, visual information, is critical. 


As visual information is crucial for ASR in certain cases, there are some previous works that use visual information for ASR directly, as in \cite{https://doi.org/10.48550/arxiv.2204.13206} or for lip reading task \cite{KR2022SubwordLL}. However, even though visual information could help ASR, the performance gain is limited by the amount of training data, where each training instance should be the triplet of vision, audio, and text information. We believe ASR error correction models, which could leverage visual data and large-scale text data, will improve the performance further.

Error correction can be viewed as similar to machine translation  (MT) task where a sequence-to-sequence model like transformers \cite{Vaswani2017AttentionIA} with large amounts of high-quality data can lead to excellent results. Inspired by multi-modal MT and other related tasks like visual text correction \cite{li-etal-2022-vision,Mazaheri2018VisualTC}, in this paper, we propose a multi-modal ASR error correction method which utilizes visual information. 



The contributions of this paper are:
    (1)We present the multi-modal dataset of ASR transcripts along with images to help its correction. The datasets were obtained from two sources, how2 dataset \cite{Sanabria2018How2AL} and from the publicly available YouTube videos. Both the sources have reference transcripts annotated by humans. And the ASR transcripts were obtained from the Huggingface wav2vec model and the Google ASR API to show that the method is independent of the ASR model;
    (2)We propose two ways to utilize the visual information for ASR text correction. Firstly, a gated fusion method where the image features are concatenated with the textual embeddings, similar to previous works \cite{li-etal-2022-vision}. Secondly, we propose a prompt-based method to better utilize large-scale text data, where the captions from the images are used as prompts for ASR correction to provide more context.


\section{Method}
\label{sec:method}

\begin{table*}[]
\centering
\begin{tabular}{c|l|cccccc}

\multirow{1}{*}{ID} & \multirow{1}{*}{Models and Variants}     & \multicolumn{2}{c|}{Google ASR API}                           & \multicolumn{2}{c|}{Huggingface wav2vec}         \\  \cline{3-6}
& & WER                 & \multicolumn{1}{c|}{SER}            & WER                      & \multicolumn{1}{c|}{SER} \\ \hline
1 & Original & 36.80 &\multicolumn{1}{c|}{100}&31.94&\multicolumn{1}{c|}{97.79} \\
2  &Transformer    &    34.14                      & \multicolumn{1}{c|}{99.42}  &   22.36     &                       \multicolumn{1}{c|}{91.45}                         \\
3  & Prompt-based       &     \textbf{33.5}         & \multicolumn{1}{c|}{\textbf{98.94}}          &       \textbf{21.13}                     & \multicolumn{1}{c|}{\textbf{90.59}}                                \\
4  & Gated fusion &           36.58        &  \multicolumn{1}{c|}{99.90}   &    25.44      &     \multicolumn{1}{c|}{95.39}                                                \\
5                                                                                                 &   Transformer + Gated fusion   &       34.8                    & \multicolumn{1}{c|}{99.42}                              &            22.92                & \multicolumn{1}{c|}{92.8}                              \\ 

6  & Transformer + Gated fusion (Filter)             &         34.54        & \multicolumn{1}{c|}{99.42}          &           22.66          & \multicolumn{1}{c|}{92.51}                         \\
7  &  Prompt-based + Gated Fusion             & 34.21      & \multicolumn{1}{c|}{98.94} &     21.80                 & \multicolumn{1}{c|}{92.12}                      \\
8  &   Prompt-based + Gated Fusion (Filter)       &       33.9          & \multicolumn{1}{c|}{99.04}          & 21.49          & \multicolumn{1}{c|}{91.74}        \\ \hline
9& Random Image captions(Caption Prompt) & 34.2 & \multicolumn{1}{c|}{99.14} &22.11&\multicolumn{1}{c|}{91.16} \\ \hline
10 & Original(Random test set) & 34.59&\multicolumn{1}{c|}{99.56}  & 30.70&\multicolumn{1}{c|}{97.6} \\
11  &Transformer(Random test set)    &    32.49    &                      \multicolumn{1}{c|}{98.68 }    &   21.48     &                       \multicolumn{1}{c|}{90.4}                  \\
12  & Prompt-based  (Random test set)             &         \textbf{30.79}             & \multicolumn{1}{c|}{\textbf{97.8}}    &       \textbf{20.6}                     & \multicolumn{1}{c|}{\textbf{89.10}}  

\end{tabular}
\caption{Measurement of error correction performance}
\label{tab:main}
\end{table*}

\vspace{-1em}
\subsection{Prompt Based Method}
\vspace{-0.5em}
Prompt-based methods have gained wide recognition following the success in several NLP tasks \cite{LiYL022}.
In this method, we use the caption data of the images as prompt for the ASR text correction task in order to provide more context. The caption data is obtained from the Flamingo model \cite{https://doi.org/10.48550/arxiv.2204.14198} trained for image captioning on the Google's Conceptual Captions datasets\footnote{\url{https://github.com/dhansmair/flamingo-mini}}. For example, consider Fig.\ref{fig:caption}b), here the caption data, "the tortoise is being cared for", is sent as a prompt by modifying the source sentence as, "the tortoise is being cared for  [SEP] thats really how we choose our tourist it for today". The modified source sentence and the target sentence are used as the parallel data for training and the modified source sentence is also used as the input while testing. Because of the presence of the caption data as prompt, more context is available to the model to correct \texttt{tourist} to \texttt{tortoise}, which otherwise would be difficult for humans too.
\vspace{-1em}
\subsection{Gated Fusion Method}
\vspace{-0.5em}
Gated fusion techniques are widely used in combining the representations from different modalities as is done in some of the previous works \cite{li-etal-2022-vision}. 
In this method, for any input sample which consists of image $I$, source text $S$ and target text $T$, the image features are obtained using the OpenAI CLIP's Vision Transformer(ViT) model \cite{Dosovitskiy2021AnII} as ViT($I$) and the textual embeddings are obtained from the standard transformer encoder as $H^S$. These two representations are fused by a vector concatenation noted as :
\begin{equation}
    H^{\mathrm{fused}} = g([H^{\mathrm{S}} ; \textbf{H}^{\mathrm{I} } ])
\end{equation}
where $H^I \in R^{L \times D}$ is the projected form of ViT($I$) to that of the length of the text representation $H^S \in R^{L \times D}$ using a linear projection layer. [L - sequence length; D - hidden dimension]. The fused representation $H^\mathrm{fused} \in R^{L \times D}$ is then passed through a \texttt{tanh} gate to control the amount of visual information used as :
\begin{equation}\label{gate}
 \Lambda = \tanh(f( [H^{\mathrm{S}} ; H^{\mathrm{fused}}  ]))
\end{equation}
The gated fused information is then added to the original textual embeddings to get the multimodal fusion representation as :
\begin{equation}\label{main}
H^{\mathrm{out}} = H^{\mathrm{S}} + \Lambda H^{\mathrm{fused}}
\end{equation}
Here tanh gate is used instead of sigmoid since it is centered at zero. This means that when $H^{\mathrm{fused}}$, is close to zero, the output of $\lambda$ will be close to zero, which aligns naturally with the situation when the image is absent as in the synthetic data. Although we found that the results were very similar for both tanh and sigmoid gates. \\
Since the high-quality annotated set with images is of a smaller size(185k samples) to correct grammatical errors, this method is applied sequentially after first obtaining the result from the baseline transformers model(Transformers+Gated fusion method in Table \ref{tab:main}). Finally, the changes made by the gated fusion method to the output from the baseline method (section \ref{sec:baseline}), is filtered by including only those changes where the similarity probability(calculated by CLIP's ViT) between the image and the changed text is higher than that with the original text(i.e output from the baseline method)[referred to as Transformers+Gated fusion(Filter) in Table \ref{tab:main}]. This is done to retain only image based corrections, since the grammar corrections are already done by the baseline method. Similar filtering and sequential correction is also applied after the prompt-based method [referred to as Prompt-based + Gated Fusion and Prompt-based + Gated Fusion(Filter) respectively in Table \ref{tab:main}].
\vspace{-0.5em}
\subsection{Dataset Retrieval}
\vspace{-0.5em}
\label{subsec:dataset}
The datasets were obtained from mainly two sources, the how2 dataset and the youtube videos. The how2 dataset consists of 300h of videos with annotated transcripts in English which resulted in 220,000 samples. The youtube videos were collected using the Youtube-DL toolkit\footnote{\url{https://github.com/ytdl-org/youtube-dl/}}, where each of these videos had annotated transcripts and audio in English. The youtube videos accounted for 2.5 million samples of data with annotated transcripts. The images for these samples were obtained by capturing the frame of the video at exactly the mid of the start and end timestamps of that sample. To obtain high-quality annotated dataset from how2 and youtube videos for visual ASR correction, we filtered the dataset to only include the examples where the similarity between the caption data and the reference transcript was greater than 0.2 as measured by Huggingface's sentence-transformers. Similarity score of 0.2 was chosen after testing performance for other thresholds\ref{tab:threshold}.  After this filtering, 58k samples were obtained from how2 dataset(accounting for 26\% of the total dataset) and 127k samples from the youtube videos(accounting for 5\% of the total dataset), indicating that a significant portion of the dataset could be corrected using the visual information. Totally, this high-quality annotated dataset consists of 185k samples with audio, images and ASR correction parallel data.

\begin{table}[h]
    \centering
    \begin{tabular}{c|c|cc}
        {Similarity Score} &{Size of train set} &\multicolumn{2}{c}{WER\hspace{0.5em} SER}   \\
        \hline
         0.1 & 380441 & 33.94&98.46\\ 
         0.2 & 183306 &\textbf{33.5} & \textbf{98.94}\\
         0.3 & 85181 & 33.89&99.23\\
    \end{tabular}
    \caption{Prompt-based method across similarity thresholds}
    \label{tab:threshold}
\end{table}





\begin{table*}[!htb]
    \centering
    \begin{tabular}{ccl}
    \toprule
      \multirow{4}{*}{\textbf{Ex.1}} & \textbf{Source sentence} & to the best strings for whatever \textbf{gucci} youre going to restring what i have on here\\
        &\textbf{Baseline} & to the best strings for whatever \textbf{gucci} youre going to restring what i have on here \\
        &\textbf{Caption prompt} & to the best strings for whatever \textbf{guitar} player youre going to restring what i have on here\\ 
        &\textbf{Target sentence} & to the best strings for whatever the \textbf{guitar} youre going to restring what i have on here \\\hline
        \multirow{4}{*}{\textbf{Ex.2}}&\textbf{Source sentence} & boxing and yeah it does because boxing kickboxing rg condola training methods\\
        &\textbf{Baseline} & boxing and yeah it does because boxing kickboxing are condola training methods \\
        &\textbf{Caption prompt} & boxing and yeah it does because boxing kickboxing are condola training methods\\ 
        &\textbf{Target sentence} & boxing and ya it does because boxing kick boxing are \textbf{jeet kune do} training methods  \\\hline
       
    \end{tabular}
    \caption{Sample Analysis}
    \label{tab:example}
\end{table*}
The training set consists of this high-quality annotated dataset along with synthetic EC parallel data. This synthetic dataset is obtained from 3 sources : 1) \textbf{TED-LIUM3} - The TED-LIUM3 corpus \cite{Hernandez2018TEDLIUM3T} is built from the TED videos; 2) \textbf{DATA2} - The DATA2 corpus \cite{Yadav2020EndtoendNE} is built for the named-entity recognition tasks; 3) \textbf{LibriSpeech} - The LibriSpeech corpus \cite{7178964} is derived from audiobooks of the LibriVox project.

A weak checkpoint trained on the GigaSpeech small dataset \cite{Chen2021GigaSpeechAE} using Neural Speech Translation toolkit \cite{zhao-etal-2021-neurst} was used to obtain the synthetic ASR transcripts by setting the beam size in the range of 8-16. Using this method, 5 million synthetic parallel data of candidate ASR transcripts and reference transcripts were obtained. 

The test set was constructed to contain 1,041 ASR transcripts from two ASR models, Google ASR API and the Huggingface wav2vec model such that the corresponding image would help in the correction of these transcripts. 

    
\vspace{-0.5em}
\section{Experiments}
\vspace{-0.5em}
\subsection{Baseline}
\label{sec:baseline}
\vspace{-0.5em}
The baseline used for this work is the standard BART-base model trained using Fairseq\footnote{\url{https://github.com/facebookresearch/fairseq}}. This model is first pre-trained with 5 million synthetic data obtained from \textit{TED-LIUM3}, \textit{DATA2} and \textit{LibriSpeech}. Then it is fine-tuned on the high-quality annotated dataset of 185k samples which were filtered based on the similarity metric of sentence-transformers model as explained in section \ref{subsec:dataset}. GPT-2 based BPE was applied in the pre-processing stage. 
\vspace{-0.5em}
\subsection{Prompt Based Method}
\vspace{-0.5em}
For the prompt-based method, a similar setup to that of the baseline was followed. The only change from the baseline was the addition of caption data as prompts in the source sentence in all three train, valid and test sets.
\vspace{-0.5em}
\subsection{Gated Fusion Method}
\vspace{-0.5em}
This method was implemented based on \href{https://github.com/libeineu/fairseq_mmt}{Multimodal Machine Translation} where the sigmoid gate function was replaced with tanh. All the parameters were kept constant as in \cite{li-etal-2022-vision}, except for the learning rate which was changed to 0.001 and the max updates to 800,000. For evaluation, the average of last 10 checkpoints was used for more reliable results. 
\vspace{-1em}
\section{Results and Analysis}
\vspace{-0.5em}
All the experiments are conducted on two different ASR models, Google ASR API and Huggingface wav2vec to verify the generality of the methods. As can be seen from Table \ref{tab:main}, the best results are obtained from the \textbf{prompt-based method}, thus verifying that visual information helps in ASR EC. 

\vspace{-0.5em}
\subsection{Relevance of Images}
\vspace{-0.5em}
To study the relevance of image information in ASR correction, we further conduct more experiments by assigning a random image to the parallel data and seeing if it can help in the ASR correction. As we can see from Table \ref{tab:main}, including random image's caption as prompt to the transformer model, leads to a decrease in performance of WER from the prompt-based method, and is on par with the baseline transformer model. This shows that the right image of the video captured during the speech is essential to improve performance.
\vspace{-0.5em}
\subsection{Why is Caption Prompt Based Method Better?}
\vspace{-0.5em}
Because it's easier to use large amounts of synthetic data for text-only methods, the prompt-based method works better than gated fusion. In the gated fusion method it is hard to correct the grammatical errors because of the lack of a large amount of parallel data with the images. In order to emphasize why caption is better, we perform experiments where, with an increase in synthetic data(Table \ref{tab:synthetic}), the prompt-based method performs better for Google ASR text correction. This shows that a better representation of the caption is learnt with increasing synthetic data and hence it can be better used as context for ASR text correction.  
\vspace{-0.5em}
\subsection{Sample Analysis}
\vspace{-0.5em}
From Ex.1 of Table \ref{tab:example}, it is clear that with additional information from the videos, it is easy to correct \texttt{gucci} to \texttt{guitar} which otherwise would be hard to correct for the transformers model. We also note the limitations of the captioning model which leads to a decrease in the performance of ASR EC. For instance, Ex.2 of Table \ref{tab:example} shows that neither the transformer model nor the caption prompt method is able to predict "jeet kune do because the caption prompt for this particular example is "person a former professional boxer is a trainer and trainer", which do not help in predicting "jeet kune do". Instead, if we have a human annotator providing the caption as "person practicing jeet kune do", it would better help in the ASR EC. Thus, due to the limitations of the captioning model, we observe that the performance of the ASR EC can be further improved with better captioning data.
\vspace{-0.5em}
\subsection{Random Test Set}
\vspace{-0.5em}
Instead of constructing the test set to have cherry-picked examples where images are needed for ASR EC, we test our proposed methods on a test set built by sampling 1000 random sentences from the how2 dataset. From the results of Table \ref{tab:main}, we can see that the gap between the baseline and caption prompt methods increases by 2x for the google ASR API and decreases only slightly for huggingface wav2vec(with still a considerable gap of 0.88\%). This shows that the caption prompt model has the potential to improve the performance of ASR text correction for instructional videos in general.

\begin{table}[]
\centering
\begin{tabular}{c|l|cccccc}

\multirow{1}{*}{ID} & \multirow{1}{*}{Method}     & \multicolumn{2}{c|}{2M Synthetic}                           & \multicolumn{2}{c|}{5M Synthetic}         \\  \cline{3-6}
& & WER                 & \multicolumn{1}{c|}{SER}             & WER                      & \multicolumn{1}{c|}{SER} \\ \hline
1  & Prompt-based               &          33.94               & \multicolumn{1}{c|}{99.33}    &       \textbf{33.5}                     & \multicolumn{1}{c|}{\textbf{98.94}}                              \\ 
\end{tabular}
\caption{EC performance with different size of synthetic data}
\label{tab:synthetic}
\end{table}


\vspace{-0.5em}
\section{Conclusion and Future Work}
\vspace{-1em}
In this work, we have used visual information to aid ASR EC, a method that has not been previously explored. We conducted several experiments to show that visual information can help in ASR EC, which would otherwise be hard to correct for strong baseline models like transformers or even humans without the context. We have introduced simple methods like using captions as prompts, which do not need any modification to the original architecture of transformers, and it improves the WER by upto 1.2\% over the baseline methods. Although we focused on text and visual methods in this work, we believe that incorporating audio information could further enhance our results, which we aim to explore in the future.




\bibliographystyle{IEEEbib}
\bibliography{strings,refs}

\end{document}